\documentclass[aps,prl,showpacs,twocolumn]{revtex4}
\usepackage{graphicx}
\usepackage{slashbox}
\usepackage{amsmath}
\usepackage{amsfonts}
\usepackage{color}
\usepackage[normalem]{ulem}
\usepackage{bm}

\begin{document}

\title{Fractional Chern Insulators in Singular Geometries}
\author{Ai-Lei He$^{1}$, Wei-Wei Luo$^{1}$, Yi-Fei Wang$^{2}$, Chang-De Gong$^{2,1}$ }
\affiliation{$^1$National Laboratory of Solid State Microstructures and Department of Physics, Nanjing University, Nanjing 210093, China  \\$^2$Center for Statistical and Theoretical Condensed Matter Physics, and Department of Physics, Zhejiang Normal University, Jinhua 321004, China}
\date{\today}

\begin{abstract}
  The fractional quantum anomalous Hall (FQAH) states or fractional Chern insulator (FCI) states have been studied on two-dimensional (2D) flat lattices with different boundary conditions. Here, we propose the geometry-dependent FCI/FQAH states that interacting particles are bounded on 2D singular lattices with arbitrary $n$-fold rotational symmetry. Based on the generalized Pauli principle, we construct trial wave functions for the singular-lattice FCI/FQAH states with the aid of an effective projection approach, and compare them with the exact diagonalization results. High wave-function overlaps show that the singular-lattice FCI/FQAH states are certainly related to the geometric factor $\beta$.  More interestingly, we observe some exotic degeneracy sequences of edge excitations in these singular-lattice FCI/FQAH states, and provide an explanation that two branches of edge excitations mix together.
\end{abstract}

%\pacs{73.43.Cd, 05.30.Jp, 71.10.Fd, 37.10.Jk}  \maketitle
\pacs{ }  \maketitle

{\it Introduction.---}Laughlin states~\cite{Laughlin} with a simple math expression but profound implications can be used to describe the ground state (GS) of fractional quantum Hall (FQH) effect~\cite{Tsui} which shows the electron fractionalization in two-dimensional (2D) systems. Over the following decades, the quantum anomalous Hall (QAH) or Chern insulator (CI) states were proposed~\cite{Haldane} and realized in ultracold fermion systems~\cite{RealHaldane}. Meanwhile, the fractional quantum anomalous Hall (FQAH) or fractional Chern insulator (FCI) states have been studied based on the topological flat band (TFB) models~\cite{TFBs,TFBs1,TFBs2} with various boundary conditions, like the torus, cylinder and disk geometries~\cite{Sheng1,YFWang1,Regnault1,YFWang2,Bernevig,FCI_reviews,FCI_reviews1,Qi0,Qi1,Qi2,Qi3,Qi4,YLWu,YLWu0,Scaffidi,WWLuo,ALHe}.  One of the most palpable features for the FCI/FQAH states is the edge excitations which are characterized by the chiral Luttinger liquid theory~\cite{XGWen}. The edge excitation spectra have been directly or indirectly observed via different numerical techniques~\cite{WWLuo,ZLiu}. These edge excitation spectra have been predicted and approximately obtained ~\cite{WWLuo,ALHe} base on the generalized Pauli principle (GPP)~\cite{GPP0,GPP1,GPP2,GPP3}. In terms of the one-to-one mapping relationship between the FQH and FCI/FQAH states, there are various ways to construct the optimal trial wave functions (WFs) for the FCI/FQAH states with analytical, semi-analytical~\cite{Qi0,Qi1,YLWu,YLWu0} and purely numerical approaches~\cite{ALHe}. Recently, inspired by the analytic expression of the Laughlin WF, we have proposed a direct yet effective prescription to construct the FCI/FQAH states on flat disk geometry~\cite{ALHe} with the aid of the GPP and the Jack polynomials (Jacks)~\cite{Jacks0,Jacks1,Jacks2}.

Recently, particularly intriguing cases occur when the topological states are proposed on some two-dimensional (2D) surfaces and lattices with a non-zero curvature, such as core states, excessive or fractional charges around the non-zero curvature parts, and multiple branches of edge excitations~\cite{Son,Ruegg,TuHH,TCan,Gromv,ALHe1}. Especially, the WFs for FQH states on the conical surface have been constructed~\cite{TuHH} in terms of the form of single-particle states with a uniform magnetic field through the cone~\cite{Bueno,Son,TuHH}. More importantly, based on Haldane's proposals, there is an implicit geometric degree of freedom for the FQH states. The original Laughlin WF is one member of the Laughlin states which are described by the spatial metric field~\cite{Haldane1,Haldane2}. The WFs for Laughlin states on the conical surface~\cite{TuHH} are similar to the Laughlin WF on disk only substituting a complex coordinate $z^{\beta}$ for $z$ where ${\beta}$ is the remaining part of disk (we call it the geometric factor and it reflects the geometric structure)~\cite{ALHe1}. The Hall viscosity and gravitational response of quantum Hall (QH) states are possible to be measured on singular geometries~\cite{Viscosity,Viscosity1,Haldane3}. Landau levels have been synthesized with photons confined on a conical surface in a recent experiment~\cite{Schine} which is expected to study the photonic FQH states~\cite{Umucalilar} and even measure the gravitational responses directly.

In this paper, we propose the FCI/FQAH states of hard-core bosons filling singular lattices with arbitrary $n$-fold rotational symmetries. We generalize the WFs for the FQH states with geometric factors ${\beta}>0$. Based on the mapping relationship between the FQH and the FCI/FQAH states and the GPP, we construct trial WFs for the singular-lattice FCI/FQAH states with the aid of an effective projection approach. To show the feasibility of our approach to construct trial WFs on singular lattices, we compare them with the exact diagonalization (ED) results. High WF overlaps confirm the reliability of these trial WFs. Meanwhile, our trial WFs pinpoint the geometrical degree of freedom of the FCI/FQAH states which are related to the geometric factor ${\beta}$. Interestingly, there are two branches of chiral edge excitation spectra observed on singular lattices with a proper trap potential. Some exotic degeneracy sequences of edge excitations in these 1/2 FCI/FQAH states are observed, and explained that two branches of edge excitations mix together. Nevertheless, these edge excitation degeneracy sequences are ruled by the GPP which has been verified in both bosonic and fermionic FCI/FQAH systems~\cite{WWLuo,ALHe}.

\begin{figure}[!htb]
\includegraphics[scale=0.450]{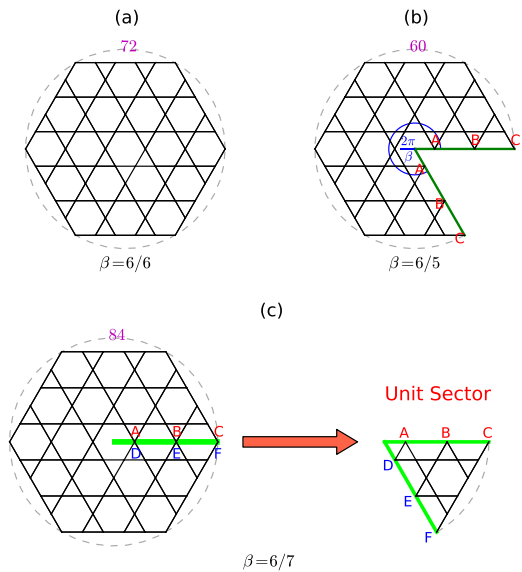}
\caption{(color online).  The kagom\'{e} lattice on planar and singular surfaces. (a) the kagom\'{e} lattice with the $6$-fold rotational symmetry on disk. (b) a $5$-fold rotational symmetric kagom\'{e} lattice can be viewed as on the conical surface with ``cutting and gluing'' point A, B and C. (c) the unfolded drawing of the $7$-fold rotational symmetric kagom\'{e} lattice is viewed as a disk geometry gluing a unit sector with the bounding point A, B, C, D, E, F. $\beta$ is related to the $n$-fold rotational symmetry. The total lattice sizes are indicated by the circles and the labelled numbers.}
\label{disk_Lattice}
\end{figure}

{\it Lattices and models.---}The kagom{\'e} lattice on the flat disk geometry with 6-fold rotational symmetry exhibits an edge [shown in Fig.~\ref{disk_Lattice}(a)]. When a sector was cut off from the kagom{\'e}-lattice disk and two corresponding cutting edges were glued together, the lattice becomes a cone. Its unfolded drawing is shown in Fig.~\ref{disk_Lattice}(b) with 5-fold rotational symmetry. Following the ``cutting and gluing'' method, the lattice with 7-fold rotational symmetry [shown in Fig.~\ref{disk_Lattice}(c)] can be defined when cutting along the radius, inserting and gluing a unit sector with $\pi/3$ angle. Based on the disk geometry and the defined unit sector, other singular geometries with arbitrary $n$-fold rotational symmetries can be constructed through the similar operations of ``cutting and gluing''~\cite{ALHe1}. In order to distinguish different lattices, we mark such a singular lattice by an angle $2\pi/\beta$ [shown in Fig.~\ref{disk_Lattice}(b)]  in which $\beta$ is related to the $n$-fold rotational symmetry, i.e. $\beta=6/n$. The flat disk geometry has $\beta=1$, the cone geometry has $\beta>1$, and $0<\beta<1$ for helicoid-like surfaces. There are some peculiar properties for the singular-lattice CI states, such as the core states, the fractional charge, and multiple branches of edge excitations~\cite{ALHe1}.

The many-body Hamiltonian for the kagom\'{e} lattice loaded with hard-core bosons ~\cite{YFWang1,WWLuo} is given by:
\begin{eqnarray}
H_{\rm KG}= &-&t\sum_{\langle\mathbf{r}\mathbf{r}^{ \prime}\rangle}
\left[b^{\dagger}_{\mathbf{r}^{ \prime}}b_{\mathbf{r}}\exp\left(i\phi_{\mathbf{r}^{ \prime}\mathbf{r}}\right)+\textrm{H.c.}\right]\nonumber\\
&-&t^{\prime}\sum_{\langle\langle\mathbf{r}\mathbf{r}^{\prime}\rangle\rangle}
\left[b^{\dagger}_{\mathbf{r}^{\prime}}b_{\mathbf{r}}+\textrm{H.c.}\right]%\nonumber
%\\&&+V_1\sum_{\langle\mathbf{r}\mathbf{r}^{ %\prime}\rangle}n_{\mathbf{r}}n_{\mathbf{r}^{\prime}}
%+V_2\sum_{\langle\langle\mathbf{r}\mathbf{r}^{
%\prime}\rangle\rangle}n_{\mathbf{r}}n_{\mathbf{r}^{\prime}}
\label{e.2}
\end{eqnarray}
Where $b^{\dagger}_{\mathbf{r}}$($b_{\mathbf{r}}$) is the hard-core boson creation (annihilation) operator at lattice site $\mathbf{r}$, and $\langle\dots\rangle$, $\langle\langle\dots\rangle\rangle$ denote the near neighbor (NN) and the next near neighbor (NNN) pairs of sites. $\phi$ is the parameter for the staggered-flux phases. Here, we choose the flat band parameters of the kagom\'{e}-lattice model, $t=1$, $t^{\prime}=-0.19$, and $\phi=0.22\pi$~\cite{WWLuo,ALHe}, to explore the FCI/FQAH states in singular geometries. The harmonic trap is added on every singular lattice to constraint bosons by the trap potential $V_{\rm trap}\sum_{\mathbf{r}} |\mathbf{r}|^2 n_{\mathbf{r}}$ with $V_{\rm trap}$ as the potential strength (with the NN hopping $t$ as the energy unit), and $|{\mathbf{r}}|$ as the radius from the center of singular lattices (with the half lattice constant $a/2$ as the length unit)~\cite{WWLuo,ALHe,ALHe1}.

{\it FQH states on singular surfaces.---}The single-particle wave function for the lowest Landau level (LLL) with a uniform magnetic field through the conical surface~\cite{Son,TuHH} is:
\begin{equation}\label{single_one}
%\phi^{\uppercase\expandafter{\romannumeral1}}_{0,m}(\{z\})=\mathcal{N}_{0,m} (z^{\beta})^m {\rm{exp}}({-|z|^2/4}),
\phi_{m}(z,\beta)=\mathcal{N}_{m} (z^{\beta})^m {\rm{exp}}({-|z|^2/4}),
\end{equation}
where $z=x+iy$ is the complex coordinate, $\beta$ marks the remaining part of a disk after cutting, $m$ marks the angular momentum quantum number in the LLL and $\mathcal{N}_{m}=\sqrt{\frac{\beta}{2\pi2^{\beta m}\Gamma(\beta m+1)}}$ is the normalization factor which is related to the geometric factor $\beta$ ~\cite{supplementary material} with the Gamma function $\Gamma(m+1)=m!$. Inspired by Ref.~\cite{TuHH}, where the $\nu=1/2$ bosonic FQH state on a cone is
\begin{equation}\label{Bose_Laughlin}
\Psi_{\rm{FQH}}(\{z_i\},\beta)=\displaystyle\prod_{i<j}(z^{\beta}_i-z^{\beta}_j)^{2}{\rm{exp}}({-\displaystyle\sum_i|z_i|^2/4}),
\end{equation}
we conjecture that many-body WFs for the FQH states in generic singular geometries, i.e. the cone, the disk and helicoid-like geometries have the invariant forms with an extended interval of $\beta$ values, i.e. $\beta>0$.

We drop the Gaussian factor ${\rm{exp}}({-\sum_i|z_i|^2/4})$ and investigate the universal polynomial structures of $\Psi_{\rm{FQH}}(\{z_i\},\beta)\propto \prod_{i<j}(z^{\beta}_i-z^{\beta}_j)^{2} $.  We expand the universal polynomial structures into the monomials which are constructed with the aid of single-particle states for the LLL marked with different angular momentum quantum numbers $m$'s~\cite{ALHe}. Here, we expand the $1/2$ Laughlin state on the singular surfaces with 3 bosons like $\Psi_{\rm FQH}^{3b}(\{z_i\},\beta)\propto(z^{\beta}_1-z^{\beta}_2)^2(z^{\beta}_1-z^{\beta}_3)^2(z^{\beta}_2-z^{\beta}_3)^2$ ~\cite{supplementary material}. We substitute the $\Psi_{[4,2,0]}$ for the symmetric monomial $({z^{\beta}_1})^{4}({ z^{\beta}_2})^{2}({z^{\beta}_3})^{0}+({z^{\beta}_1})^{4}({z^{\beta}_2})^{0}({z^{\beta}_3})^{2}+({ z^{\beta}_1})^{2}({ z^{\beta}_2})^{4}({ z^{\beta}_3})^{0}+({z^{\beta}_1})^{2}({z^{\beta}_2})^{0}({z^{\beta}_3})^{4}+({z^{\beta}_1})^{0}
({z^{\beta}_2})^{4}({ z^{\beta}_3})^{2}+({ z^{\beta}_1})^{0}({ z^{\beta}_2})^{2}({ z^{\beta}_3})^{4}$.  Following this simplification method~\cite{ALHe}, the 1/2 bosonic FQH state on singular surfaces is written as $\Psi_{\rm{FQH}}^{3b}(\{z_i\},\beta)\propto(+1) \Psi_{[4,2,0]}+(-2) \Psi_{[4,1,1]}+ (-2)\Psi_{[3,3,0]}+(+2)\Psi_{[3,2,1]}+(-6)\Psi_{[2,2,2]}$.
That is to say, the symmetric monomials constitute a set of basis functions with the same total angular momentum and the bosonic FQH states can be expanded with the aid of these symmetric monomials.

The Jacks provide a simple and convenient approach to obtain the expansions of the $1/2$ FQH state with more particles, i.e. $\Psi_{\rm{FQH}} (\{z_i\})\propto \sum_{\mu \le \lambda} b_{\lambda \mu} \Psi_{\mu}$.
Here $\lambda$ is the ``root configuration'' of the basis states like $\lambda$=[$k0^{r-1}k0^{r-1}...k$], and $b_{\lambda \mu}$ is the expansion coefficient. $\Psi_{\mu}$ denotes the monomial and $\mu$ stores the angular momentum quantum numbers of every partition~\cite{Jacks0,Jacks1,Jacks2}. The Jacks can be viewed as one of the
descriptions for the GPP~\cite{GPP0,GPP1,GPP2,GPP3} in Fock space for the FQH states. The $1/2$ bosonic Laughlin state can be explained that there is no more than one particle occupying any two adjacent orbitals~\cite{Jacks0,Jacks1,Jacks2}. For the FQH states $\Psi_{\rm{FQH}}(\{z_i\},\beta)$ on singular surfaces, expansion coefficients of every symmetric monomial are the same as the FQH states in disk geometry which are obtained by the recurrence relation~\cite{Jacks0,Jacks1,Jacks2}. The normalization factor of these FQH states are related to the geometric factor $\beta$ after a simple calculation~\cite{supplementary material}. This indicates the geometric factor plays a pivotal role in the singular-surface FQH states. We use the root configuration $|1010101...101\rangle_{\rm FQH}$ to denote the $1/2$ bosonic FQH state. The root configuration of the first excited state is $|1010101...1001\rangle_{\rm FQH}$.

{\it Construction of trial wave functions.---}There is a one-to-one mapping relationship between the FQH states and the FCI/FQAH states~\cite{Qi0,Qi1,Qi2,Qi3,Qi4,YLWu,YLWu0,ALHe}. Following our previous work~\cite{ALHe}, the WFs for $1/3$ fermionic FCI/FQAH states in disk geometry can be constructed based on the GPP, the Jacks and the single-particle states with the general form, $\Psi_{\rm{FCI}} ^{1/3}(\{z_i\})=\sum_{i} c_{i} \Phi_{i}$, where $c_{i}$ is the expansion coefficient of the Jacks which is related to the angular momentum quantum number, and $\Phi_{i}$ is antisymmetric Slater determinant composed by the single-particle states of CIs.

\begin{figure}[!htb]
\includegraphics[scale=0.50]{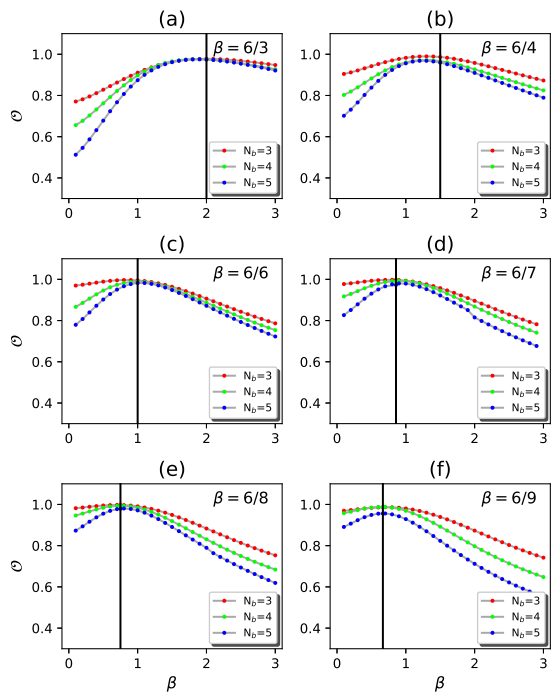}
\caption{(color online). Overlaps between the trial ground-state WFs and the ED results for the FCI/FQAH states on singular lattices and disk geometry. We construct the trial WFs based on the GPP and the Jacks. Based on the trial FQH WFs on the cone, in (a)-(f), we compare the trial WFs for the FCI/FQAH states with variable geometric factors $\beta$'s and the ED results on the singular lattices with $3$-, $4$-, $6$-, $7$-, $8$-, $9$-fold rotational symmetries. The vertical line denotes the inherent geometric factor $\beta$ of each lattice.}
\label{overlap0}
\end{figure}

For the $1/2$ bosonic FCI/FQAH state in disk or singular geometries filled with hard-core bosons in the TFBs, we assume that the trial WFs are obtained based on the GPP and the Jacks. With the aid of the mapping relationship,  we first construct the soft-core bosonic (SCB) WFs for the FCI/FQAH states $\Psi_{\rm SCB}$ in disk or singular geometries only by substituting the single-particle states $\{\psi^{i}_{\rm CI}\}$ of CIs in the TFBs for the WFs $\{\phi_{m}(z,\beta)\}$ with angular momentum quantum number $m$ in the LLL ~\cite{supplementary material}. The basis of bosonic Jacks can not satisfy the property of hard-core bosons (HCB) which exist the states with more than one particle occupying in the same lattice site. In order to reflect the HCB peculiarity, we define the projection operator which projects the SCB states into the Hilbert space of HCB states as follows,
\begin{equation}\label{projection}
\Psi_{\rm HCB}={\cal N}_{\rm FCI} \widehat{\cal P} \Psi_{\rm SCB}={\cal N}_{\rm FCI}\left[\prod_{i,j}(1-{\delta_{ij}})\Psi_{\rm SCB}\right].
\end{equation}
Here, $\Psi_{\rm HCB}$ denotes the trial WF of HCB state, $\widehat{\cal P}$ is the projection operator and $\widehat{\cal P}=\prod_{i,j}(1-{\delta_{ij}})$ which means no more than one particles occupying in the same lattice site with $i,j$ marking the site positions and traversing all lattice sites. $\delta_{ij}$ is the Kronecker delta function.
The trial WFs for the singular-lattice FCI/FQAH states are obtained after normalizing the projection WFs with the normalization factor ${\cal N}_{\rm FCI}$. In fact, for the FCI/FQAH states of soft-core bosons, the weight of more than one bosons occupying in the same sites is very small ~\cite{supplementary material}.

\begin{figure}[!htb]
\includegraphics[scale=0.50]{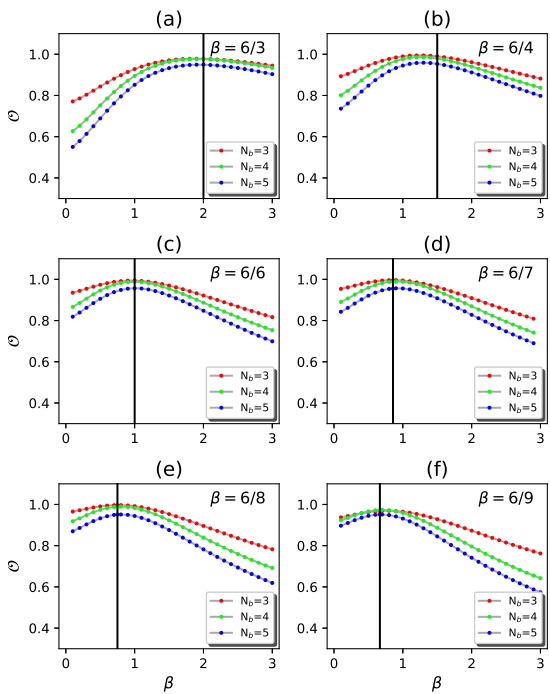}
\caption{(color online).  Overlaps between the first excited-state WFs and the ED results for the FCI/FQAH states on singular lattices and disk geometry. In (a)-(f), we compare the trial WFs for the FCI/FQAH states with variable geometric factor $\beta$ and the ED results on the singular lattices with $3$-, $4$-, $6$-, $7$-, $8$-, $9$-fold rotational symmetries. The inherent geometric factor $\beta$ of each lattice is marked by the vertical line.}
\label{overlap1}
\end{figure}

{\it Wave-function overlaps.---}In order to investigate the reliability of the approach to construct the trial WFs, we compare the trial WFs with the ED results, i.e. calculating the overlap values ${\cal{O}}=|\langle \Psi_{\rm HCB}|\Psi_{\rm ED}\rangle|$~\cite{ALHe}. According to the analysis of the FQH states on conical surfaces, we conjecture that singular-lattice FCI/FQAH states are related to the geometric factor $\beta$. In order to confirm our conjecture, a series of trial WFs are constructed with the continuously variable geometric factor $\beta$.  We consider the $1/2$ FCI/FQAH states (the root configuration $|1010101...101\rangle_{\rm FCI}$) on the singular lattices with 3-, 4-, 7-, 8- and 9-fold rotational symmetry and the disk geometry with 6-fold rotational symmetry filled with 3-, 4- and 5- hard-core bosons. Here, we show the WFs overlap values with the continuously variable geometric factor $\beta$ and we specifically mark the values of every singular lattice geometric factor $\beta$ with black line in Fig.~\ref{overlap0}. It is obvious that overlaps between the trial WFs with the inherent geometric factor $\beta$ and the ED results are in the vicinity of the maximum. High values of WFs overlap show the singular-lattice FCI/FQAH states are certainly related to the geometric factor $\beta$. Analogously, the first excited WFs for the FCI/FQH states with the root configuration $|1010101...1001\rangle_{\rm FCI}$ can be constructed based on the GPP and the Jacks. Undoubtedly, these WFs are related to the geometric factor $\beta$ from the WFs overlap values in Fig.~\ref{overlap1}. Values of the WFs overlap are quite high(more than 0.95 even for the FCI/FQAH states filled with 5 hard-core bosons), however, the Hilbert-space dimensions of the Jacks are quite fewer than the ED dimension~\cite{supplementary material} with the suitable geometric factor $\beta$.

\begin{figure}[!htb]
\includegraphics[scale=0.50]{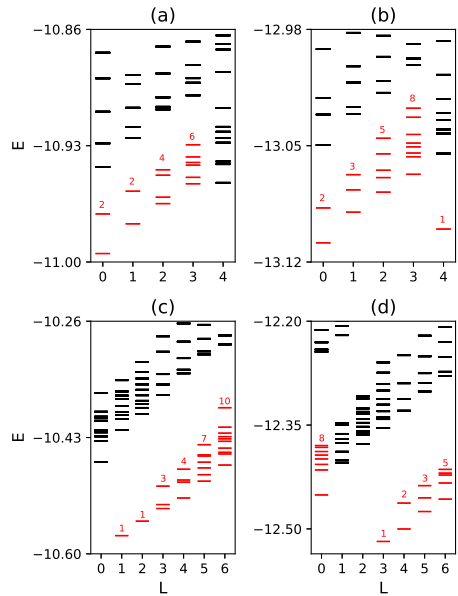}
\caption{(color online). Edge excitations for $1/2$ FCI/FQAH states on singular lattices. In (a)-(b), 5 and 6 hard-core bosons fill in the 5-fold rotational symmetric lattice with the trap $V_{\rm trap}=0.0025$ and in (c)-(d), 5 and 6 hard-core bosons fill in 7-fold rotational symmetric lattice with the trap $V_{\rm trap}=0.008$. The edge excitation sequence is partly marked by the red symbols and numbers.}
\label{Edge_Excitation}
\end{figure}

{\it Edge excitations.---}Edge excitations for FCI/FQAH states on a flat disk geometry have been directly obtained based on the numerical approach and the quasi-degeneracy sequence in low-energy edge excitations is predicted based on the GPP~\cite{WWLuo,ALHe}. There is only one branch of edge excitations for the $1/2$ bosonic or $1/3$ fermionic FCI/FQAH states in flat disk which is quite coincident with the degeneracy sequences ``1,1,2,3,5,7,...''. However, there are more than one branches of edge excitations for the singular-lattice CI/QAH states in which the the number of energy branches is related to the number of the core states~\cite{ALHe1}. Similarly, there are more than one branches of edge excitations for the $1/2$ bosonic FCI/FQAH states with the core states on the singular lattice.  Interestingly, we obtain the edge excitations spectra for these FCI/FQAH states with $5$- and $7$-fold rotational symmetry (shown in Figure~\ref{Edge_Excitation}). Evidently, the degeneracy sequences of the edge excitations are not ``1,1,2,3,5,7,...''. In fact, these exotic edge excitation sequences are explained that two branches (with one core state) of edge excitations mix together and each branch of edge excitations fulfills the GPP ~\cite{supplementary material}. And these exotic edge excitations are also different from the 2/5 and 2/3 FQH edge states which consist of two independent droplets~\cite{XGWen,XGWen1} instead of the core states~\cite{ALHe1}.

{\it Summary.---}We propose the FCI/FQAH states on singular lattices which are marked with the $n$-fold ($n=3,4,5$...) rotational symmetries or the geometric factor $\beta=6/n$. We use the numerical ED method to obtain the low-energy many-body states for the singular-lattice FCI/FQAH states. Based on the GPP and the Jacks, we put forward the projection method and succeed in constructing the trial WFs for the singular-lattice FCI/FQAH states of hard-core bosons. Our trial WFs are related to the geometric factor $\beta$ and the high WF overlaps with the numerical ED WFs reveal the significance of the geometrical degree of freedom in the FCI/FQAH states. More interestingly, due to the core states in singular geometries, exotic degeneracy sequences of the edge excitations for these singular-lattice FCI/FQAH states have been observed and explained.

{\it Acknowledgements.---}This work is supported by the NSFC of China Grants No.11874325, No.11374265 (Y.F.W.), and the State Key Program for Basic Researches of China Grant No.2009CB929504 (C.D.G.).

\clearpage
%\newpage

\section*{Supplementary Material for ``Fractional Chern Insulators in Singular Geometries''}
In the main text of this paper, we construct the trial WFs and obtain two branches of edge excitations for the $1/2$ FCI/FQAH states of hard-core bosons on singular lattices with different folds of rotational symmetry based on the GPP and the Jacks. Here, we'll show some details to construct the trial WFs including the normalization constant of the Jacks with geometric factor $\beta$, the one-to-one mapping relationship and how to project the soft-core bosonic (SCB) states into the hard-core bosonic (HCB) Hilbert spaces. Because of the core state around the defect lattice, there are two branch of edge excitations and these novel edge excitations also fulfil the GPP. Here, we use some diagrammatic illustration to describe how to generate the root configurations of edge excitations with core states.

\section*{Expansion, Normalization and geometric factor $\beta$}
The $1/2$ FQH states on the cone is
\begin{equation}\label{FQH_cone}
\Psi_{\rm{FQH}}(\{z_i\},\beta)=\displaystyle\prod_{i<j}(z^{\beta}_i-z^{\beta}_j)^{2}{\rm{exp}}({-\displaystyle\sum_i|z_i|^2/4})
\end{equation}
with the geometric factor $\beta$ and here, we show the procedure of obtaining the normalization constant based on the Jacks. We first take the FQH states with 2 bosons as an example. The Jacks expansion is
\begin{eqnarray}
&&\Psi_{\rm FQH}^{2b}(\{z_i\},\beta)=(z^{\beta}_1-z^{\beta}_2)^2\exp\left[-(|z_1|^2+|z_2|^2)/4\right]\nonumber\\
&&=\left\{(+1)\left[({z^{\beta}_1})^{2}({ z^{\beta}_2})^{0}+({z^{\beta}_1})^{0}({ z^{\beta}_2})^{2}\right]+(-2)({z^{\beta}_1})^{1}({ z^{\beta}_2})^{1}\right\}\nonumber\\&& \exp\left[-(|z_1|^2+|z_2|^2)/4\right]\nonumber\\&&=(+1) \mathcal{N}_{[2,0]}
\left[\phi_{2}(z_1,\beta)\phi_{0}(z_2,\beta)+\phi_{0}(z_1,\beta)\phi_{2}(z_2,\beta)\right]\nonumber\\&&+(-2) \mathcal{N}_{[1,1]} \phi_{1}(z_1,\beta) \phi_{1}(z_2,\beta).
\label{Expand_polynomial1}
\end{eqnarray}
Here, $\phi_{m}(z,\beta)$ is the single-particle state in the LLL with angular momentum quantum number $m$. $\mathcal{N}_{[2,0]} =\sqrt{\frac{(2\pi)^2(2^\beta)^{2+0}\Gamma(2\beta+1)\Gamma(1)}{(\beta)^2}}$ and $\mathcal{N}_{[1,1]}=\sqrt{\frac{(2\pi)^2(2^\beta)^{1+1}(\Gamma(\beta+1))^2}{(\beta)^2}}$ are related to the normalization constant of the single-particle state in eq.~\ref{single_one}.  Let's calculate the ${\cal O}^{\rm 2b}=\langle \Psi_{\rm FQH}^{2b}(\{z_i\},\beta)| \Psi_{\rm FQH}^{2b}(\{z_i\},\beta) \rangle$, based on the orthogonal relation $\langle \phi_{m}| \phi_{n} \rangle=\delta_{m,n}$, one can obtain ${\cal O}^{\rm 2b}= {\mathcal{N}^2_{[2,0]}}+4{\mathcal{N}^2_{[1,1]}}$.
%${\cal O}^{\rm 2b}= {\mathcal{N}^2_{[2,0]}}[|\langle %\phi_{2}(z_1,\beta)|\phi_{2}(z_1,\beta)\rangle|^2|\phi_{0}(z_2,\beta)|^2+|\phi_{0}(z_1,\beta)|^2|\phi_{2}(z_2,\beta)|^2]
%+4{\mathcal{N}^2_{[1,1]}}|\phi_{1}(z_1,\beta)|^2|\phi_{1}(z_2,\beta)|^2$.
So, the normalization constant is $\frac{1}{\sqrt{{\cal O}^{\rm 2b}}}=\frac{1}{\sqrt{2{\mathcal{N}^2_{[2,0]}}+4{\mathcal{N}^2_{[1,1]}}}}$, and the normalized wave function is,
\begin{eqnarray}
&&\Psi_{\rm FQH}^{2b}(\{z_i\},\beta)=(z^{\beta}_1-z^{\beta}_2)^2\exp\left[-(|z_1|^2+|z_2|^2)/4\right]=\nonumber\\&&
\left\{\frac{(+1)}{\sqrt{{\cal O}^{\rm 2b}}}\left[({z^{\beta}_1})^{2}({ z^{\beta}_2})^{0}+({z^{\beta}_1})^{0}({ z^{\beta}_2})^{2}\right]+\frac{(-2)}{\sqrt{{\cal O}^{\rm 2b}}}({z^{\beta}_1})^{1}({ z^{\beta}_2})^{1}\right\}\nonumber\\&&\exp\left[-(|z_1|^2+|z_2|^2)/4\right]=\nonumber\\&& \frac{(+1)}{\sqrt{{\cal O}^{\rm 2b}}} {\mathcal{N}_{[2,0]}} [\phi_{2}(z_{1},\beta)\phi_{0}(z_{2},\beta)+\phi_{0}(z_{1},\beta)\phi_{2}(z_{2},\beta)]\nonumber\\&& + \frac{(-2)}{\sqrt{{\cal O}^{\rm 2b}}} {\mathcal{N}_{[1,1]}} \phi_{1}(z_1,\beta)\phi_{1}(z_2,\beta).
\label{Normal_1}
\end{eqnarray}
The simplified form of the WF for 2 bosons is $\Psi_{\rm FQH}^{2b}(\{z_i\},\beta)\propto(+1)\Psi_{[2,0]}+(-2)\Psi_{[1,1]}$ and the probability of the $\Psi_{[2,0]}$ is $\frac{2{\mathcal{N}^2_{[2,0]}}}{2{\mathcal{N}^2_{[2,0]}}+4{\mathcal{N}^2_{[1,1]}}} \propto \Gamma(2\beta+1)\Gamma(1)$ and the probability of the $\Psi_{[1,1]}$ is $\frac{4{\mathcal{N}^2_{[1,1]}}}{2{\mathcal{N}^2_{[2,0]}}+4{\mathcal{N}^2_{[1,1]}}}\propto \Gamma(\beta+1)\Gamma(\beta+1)$.

Next, we normalize the $1/2$ Laughlin state for 3 bosons on the conical surface based on the expand polynomial (Eq.~\ref{Expand_polynomial}) (without the Gaussian factor).
\begin{eqnarray}
&&\Psi_{\rm FQH}^{3b}(\{z_i\},\beta)\propto(z^{\beta}_1-z^{\beta}_2)^2(z^{\beta}_1-z^{\beta}_3)^2(z^{\beta}_2-z^{\beta}_3)^2\nonumber\\&&
=(+1)[({z^{\beta}_1})^{4}({ z^{\beta}_2})^{2}({z^{\beta}_3})^{0}+({z^{\beta}_1})^{4}({z^{\beta}_2})^{0}({z^{\beta}_3})^{2}+({ z^{\beta}_1})^{2}\nonumber\\&& ({ z^{\beta}_2})^{4}({ z^{\beta}_3})^{0} +({z^{\beta}_1})^{2}({z^{\beta}_2})^{0}({z^{\beta}_3})^{4}+({z^{\beta}_1})^{0}
({z^{\beta}_2})^{4}({ z^{\beta}_3})^{2}\nonumber\\&&+({ z^{\beta}_1})^{0}({ z^{\beta}_2})^{2}({ z^{\beta}_3})^{4}] +(-2)[({z^{\beta}_1})^{4}({z^{\beta}_2})^{1}({ z^{\beta}_3})^{1}+ ({z^{\beta}_1})^{1}\nonumber\\&& ({ z^{\beta}_2})^{4}({z^{\beta}_3})^{1}+
({ z^{\beta}_1})^{1}({z^{\beta}_2})^{1}({z^{\beta}_3} )^{4}]+(-2)[({z^{\beta}_1})^{3}({z^{\beta}_2})^{3} \nonumber\\&& ({z^{\beta}_3})^{0}+({z^{\beta}_1})^{3}({ z^{\beta}_2})^{0}({ z^{\beta}_3})^{3}+({ z^{\beta}_1})^{0}({z^{\beta}_2})^{3}({z^{\beta}_3})^{3}]
+(+2) \nonumber\\&& [({z^{\beta}_1})^{3}({ z^{\beta}_2})^{2}({ z^{\beta}_3})^{1}+({ z^{\beta}_1})^{3}({ z^{\beta}_2})^{1}({ z^{\beta}_3})^{2} +({z^{\beta}_1})^{1}({z^{\beta}_2})^{2} \nonumber\\&& ({z^{\beta}_3})^{3}+({ z^{\beta}_1})^{1}({ z^{\beta}_2})^{3}({z^{\beta}_3})^{2}
+({ z^{\beta}_1})^{2}({ z^{\beta}_2})^{1}({z^{\beta}_3})^{3}
+({z^{\beta}_1})^{2} \nonumber\\&& ({ z^{\beta}_2})^{3}({z^{\beta}_3})^{1}]+(-6){(z^{\beta}_1})^{2}({ z^{\beta}_2})^{2}({z^{\beta}_3})^{2}
\label{Expand_polynomial}
\end{eqnarray}

Following the case of 2 bosons, the normalization constant for the FQH state with 3 bosons (from the Eq.~\ref{Expand_polynomial}) is ${\cal O}^{\rm 3b}=\langle \Psi_{\rm FQH}^{3b}(\{z_i\},\beta)| \Psi_{\rm FQH}^{3b}(\{z_i\},\beta) \rangle$ and one can obtain ${\cal O}^{\rm 3b}=6{\mathcal{N}^2_{[4,2,0]}}+12{\mathcal{N}^2_{[4,1,1]}}+12{\mathcal{N}^2_{[3,3,0]}}
+24{\mathcal{N}^2_{[3,2,1]}}+36{\mathcal{N}^2_{[2,2,2]}}$ with ${\mathcal{N}^2_{[4,2,0]}}=[(2\pi)^3(2^\beta)^6\Gamma(4\beta+1)\Gamma(2\beta+1)]/{\beta^3}$, ${\mathcal{N}^2_{[4,1,1]}}=[(2\pi)^3(2^\beta)^6\Gamma(4\beta+1)\Gamma(\beta+1)\Gamma(\beta+1)]/{\beta^3}$, ${\mathcal{N}^2_{[3,3,0]}}=[(2\pi)^3(2^\beta)^6\Gamma(3\beta+1)\Gamma(3\beta+1)]/{\beta^3}$, ${\mathcal{N}^2_{[3,2,1]}}=[(2\pi)^3(2^\beta)^6\Gamma(3\beta+1)\Gamma(2\beta+1)\Gamma(\beta+1)]/{\beta^3}$ and ${\mathcal{N}^2_{[2,2,2]}}=[(2\pi)^3(2^\beta)^6\Gamma(2\beta+1)\Gamma(2\beta+1)\Gamma(2\beta+1)]/{\beta^3}$.  The simplified form is $\Psi_{\rm FQH}^{3b}(\{z^{\beta}_i\})=(+1)\Psi_{[4,2,0]}+(-2)\Psi_{[4,1,1]}+(-2)\Psi_{[3,3,0]}+(+2)\Psi_{[3,2,1]}+(-6)\Psi_{[2,2,2]}$ and the probabilities of every symmetric monomial are $\frac{6{\mathcal{N}^2_{[4,2,0]}}}{{\cal O}^{\rm 3b}}\propto \Gamma(4\beta+1)\Gamma(2\beta+1)\Gamma(1)$ for $\Psi_{[4,2,0]}$, $\frac{12{\mathcal{N}^2_{[4,1,1]}}}{{\cal O}^{\rm 3b}}\propto \Gamma(4\beta+1)\Gamma(\beta+1)\Gamma(\beta+1)$ for $\Psi_{[4,1,1]}$, $\frac{12{\mathcal{N}^2_{[3,3,0]}}}{{\cal O}^{\rm 3b}}\propto \Gamma(3\beta+1)\Gamma(3\beta+1)\Gamma(1)$ for $\Psi_{[3,3,0]}$, $ \frac{24{\mathcal{N}^2_{[3,2,1]}}}{{\cal O}^{\rm 3b}}\propto \Gamma(3\beta+1)\Gamma(2\beta+1)\Gamma(\beta+1)$ for $\Psi_{[3,2,1]}$, and $\frac{36{\mathcal{N}^2_{[2,2,2]}}}{{\cal O}^{\rm 3b}}\propto \Gamma(2\beta+1)\Gamma(2\beta+1)\Gamma(2\beta+1)$ for $\Psi_{[2,2,2]}$. After these two examples, we obtain the
probabilities of every expansion basis of the FQH states on cone based on the Jacks and they are related to the geometric factor $\beta$.

\begin{figure}[!htb]
\includegraphics[scale=0.40]{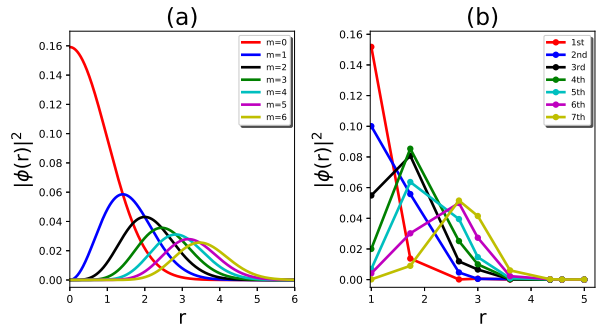}
\caption{(color online). Mapping relationship between the QH states and the CI states. (a) Single-particle states of the LLL in disk along the radial direction with various angular momentum quantum numbers $m$'s. (b) Single-particle states for some low energy levels on the kagom\'{e} lattice disk with the $C_6$ rotational symmetry. We choose the 72-site kagom{\'e} lattice with the harmonic trap $V_{\rm trap}=0.005$. Different states are marked with different colors.}
\label{single_particle_states}
\end{figure}

\section*{One-to-one mapping relationship and WF overlap}
It's obvious that the changes of these single-particle states along the radial direction have the same trends between the continuous system and CI lattice. If we replace the single-particle states in the LLL with the single-particle states of CI, the wave functions of FCI/FQAH states can be constructed. This substitutional relation reflects the one-to-one mapping relationship between FQH and FCI/FQAH states. Here, it's obvious to reflect the mapping relationship between the QH states and the CIs states with the same variation tendency shown in Fig.\ref{single_particle_states}. We can substitute the single-particle states $\{\psi^{i}_{\rm CI}\}$ of CIs in TFBs for the wave functions with angular momentum quantum number $m$ in the LLL $\{\phi_{m}(z,\beta)\}$. The mapping relationship can be reflected by these two states with the same color like the $m=0$ single-particle state in LLL and the 1st CI state, the $m=1$ single-particle state in LLL and the 2ed CI state and so on shown in Fig.\ref{single_particle_states}. Based on the mapping relationship and the Jacks, we can construct the bosonic FCI/FQAH states. The WF overlaps and the Hilbert-space dimensions of the ED results and the trial WF based on the Jacks are shown in the table~\ref{overlap_max}. However, this many-body WF (of soft-core bosons) cannot fulfill one lattice site with no more than one particles (i.e. the hard-core constraint), so in the following, we introduce how to project the soft-core bosonic WFs into the hard-core Hilbert space.

\begin{table}
\begin{tabular} {c c c c c c c c}
 \hline\hline
$N_b$~& ~$n$ ~& ~$\beta$  & ~${\cal D}_{\rm ED}$  ~& ~${\cal D}^{\rm GS}_{\rm Jack}$ ~& ~${\cal O}_{\rm GS}$~ &~${\cal D}^{\rm ES}_{\rm Jack}$~ & ~${\cal O}_{\rm ES}$ \\
\hline
4 & 4 & 6/4 & 194580 & 16 & 0.967 &20 &0.980 \\
%\hline
5 & 4& 6/4 & 1712304 & 59 & 0.959 &77& 0.955 \\
% \hline
4 & 6 & 6/6 & 111930 & 16 & 0.994 & 20& 0.988 \\
%\hline
5 &6 & 6/6 & 850668 & 59 & 0.982 & 77 & 0.966 \\
% \hline
4 & 7 & 6/7 & 595665 & 16 & 0.993 & 20& 0.989 \\
%\hline
5 &7 & 6/7 & 7028847 & 59 & 0.979 & 77 & 0.956 \\
% \hline
4 & 8 & 6/8 & 1028790 & 16 & 0.994 & 20& 0.989  \\
%\hline
5 &8 & 6/8 &13991544 & 59 & 0.979 & 77 & 0.951 \\
 \hline
\end{tabular}
\caption{Overlap of $1/2$ FCI/FQAH states [ground state (GS) and the first excited state (ES) ] on the kagom\'{e}-disk filled with $N_b$ hard-core bosons on singular lattice with the $n$-fold rotational symmetry (geometric factor $\beta=6/n$).  The maximum values of the overlap for the GS (${\cal O}_{\rm GS}$) and for the ES (${\cal O}_{\rm ES}$) are listed. We compare the dimensions between the ED and the Jacks, and ${\cal D}_{\rm ED}$, ${\cal D}^{\rm GS}_{\rm Jack}$ and ${\cal D}^{\rm ES}_{\rm Jack}$ denote the dimensions of the ED, the GS Jacks and the ES Jacks.}
\label{overlap_max}
\end{table}

\section*{The soft-core and hard-core Hilbert spaces}

We use the projection operator $\cal P$ to construct the FCI/FQAH states of hard-core bosons. However, using the single-particle states of CI to construct the FCI/FQAH states of hard-core bosons is different from the fermionic situation. In order to illustrate the difference, we consider a 72-site lattice (shown in Fig.~\ref{disk_Lattice}(a)) filled with 3 particles in disk geometry. The soft-core bosonic basis for FCI/FQAH states can be generated based on the mapping relationship and some expansion basis of the Laughlin wave function (eq.~\ref{Bose_Laughlin}). The dimension of Hilbert space for the soft-core bosonic FCI/FQAH state on kagom{\'e} disk is $C^3_{72}+72\times72=64824$, however, the dimension of Hilbert space for the hard-core bosonic FCI/FQAH state is $C^3_{72}=59640$. It's clear that the soft-core bosonic state dimension is larger than the hard-core bosonic state. Considering the 1/2 soft-core bosonic FCI/FQAH state $\Psi_{\rm SCB}$ in disk geometry, the weight of one site occupying more than one bosons can be defined as ${\cal W}=\left|1-\left[\prod_{i,j}(1-{\delta_{ij}})\Psi_{\rm SCB}\right]\right|^2/\left|\Psi_{\rm SCB}\right|^2$. In table~\ref{proportion}, we compare the FCI/FQAH states of the hard-core bosons and the soft-core bosons through the Hilbert space dimensions for many-body wave functions and list the weight ${\cal W}$. The difference of these two Hilbert space dimensions is very large, however, the weight of more bosons occupying in the same sites is very small.

\begin{table}
\begin{tabular} {c| c c c| c c c}
 \hline\hline
 & & $N_s=42$ & & & $N_s=72$ \\
 \hline
$N_b$ & ~${\cal D}_{\rm HCB}$  ~& ~${\cal D}_{\rm SCB}$ ~& ~${{\cal W}}$~ &~${\cal D}_{\rm HCB}$~ & ~${\cal D}_{\rm SCB}$ ~& ~${\cal W}$\\
\hline
2 & 861 & 903 & 0.0025 &2556 & 2628 & 0.0025\\
%\hline
3 & 11480 & 13244 & 0.0027 & 59640& 64824 & 0.0030\\
% \hline
4 & 111930 & 148995 & 0.0076 & 1028790 & 1215450 & 0.0084\\
%\hline
5 & 850668 & 1370754 & 0.0121 &13991544 & 18474840 &0.0155\\
 \hline
\end{tabular}
\caption{Hilbert space dimensions for the FCI/FQAH states of hard-core bosons (HCB) and soft-core bosons (SCB) and the proportion (${\cal W}$) of more than one bosons occupying in one site. Though the Hilbert space dimension for SCB wave functions is larger than the HCB ones, the weight of more bosons occupation in one site is slim. Here, $N_s$ and $N_b$ denote the numbers of lattice sites and bosons.}
\label{proportion}
\end{table}

\begin{figure*}
  \vspace{-0.0in}
\includegraphics[width=14cm]{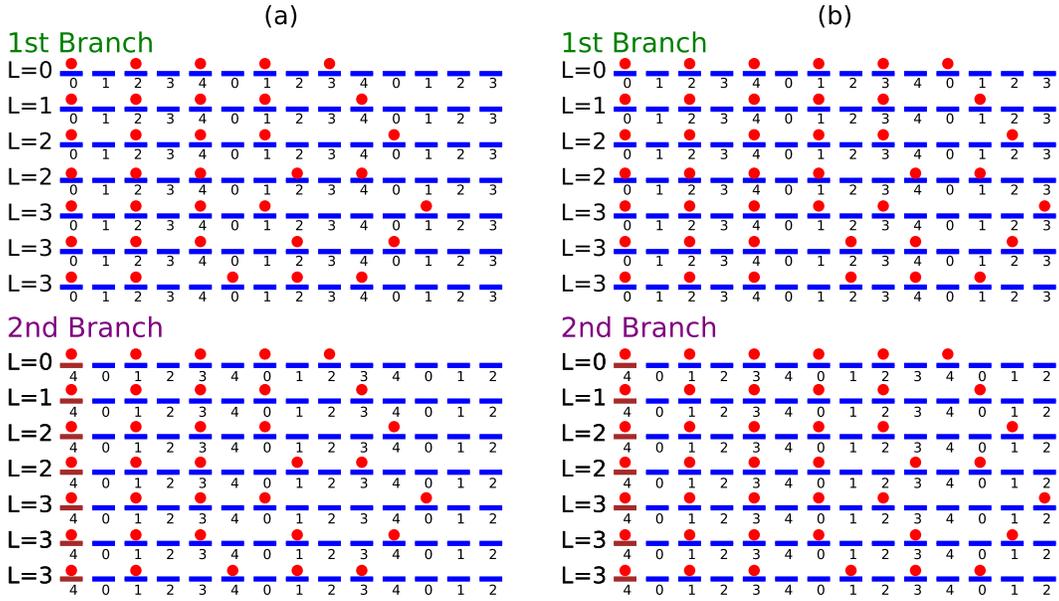}
  \vspace{-0.0in}
\caption{(color online).  Diagrammatic illustration to describe how to generate the root configurations of edge excitations with core states for the CI lattices with $5$-fold  rotational symmetry. These brown lines denote the core states and the red dots denote the filled hard-core bosons. The numbers under the short line are the angular momentum quantum numbers. L on the left in every line is the reduced angular momentum quantum number (total angular momentum quantum number modulo $5$). We only enumerate some lower-energy root configurations in every branch of edge excitations and higher-energy root configurations can be obtained based on the GPP.}
\label{Edge_GPP}
\end{figure*}

\begin{figure*}
  \vspace{-0.0in}
\includegraphics[width=14cm]{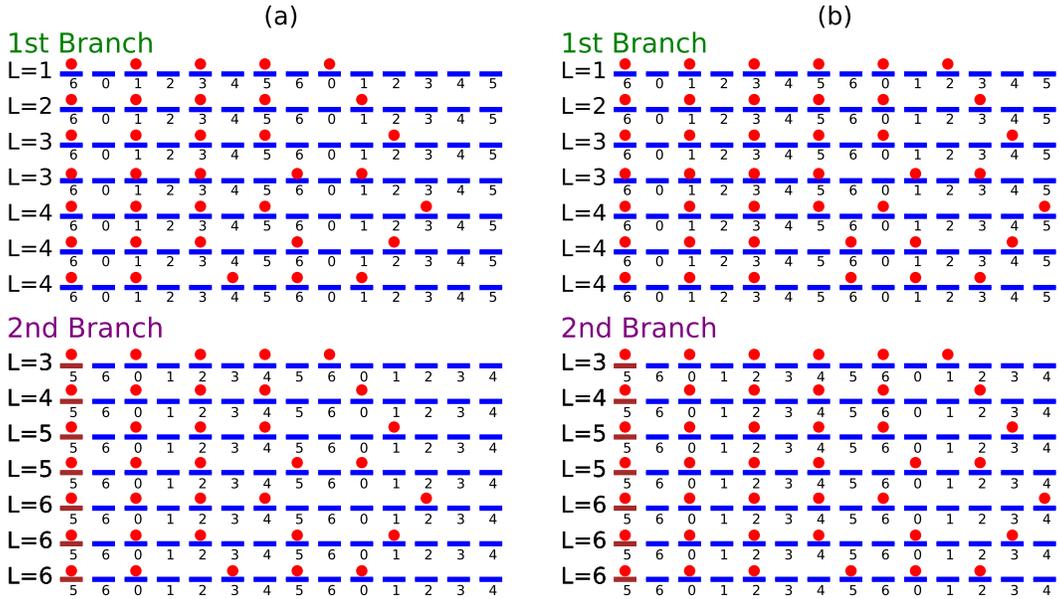}
  \vspace{-0.0in}
\caption{(color online).  Diagrammatic illustration to describe how to generate the root configurations of edge excitations with core states for the CI lattices with $7$-fold rotational symmetry. L on the left in every line is the reduced angular momentum quantum number (total angular momentum quantum number modulo $7$). We only enumerate some lower-energy root configurations in every branch of edge excitations and higher-energy root configurations can be obtained based on the GPP.}
\label{Edge_GPP1}
\end{figure*}

\section*{Edge excitations and the GPP}
In the main text, we have shown the exotic edge excitation sequences of the singular-lattice FCI/FQAH states. Here, we use a diagrammatic illustration (shown in Fig.~\ref{Edge_GPP}) to describe the occupation of hard-core bosons in more detail and tell how to write these root configurations of edge excitations easily in these singular systems based on the GPP. Here, we consider the kagom{\'e} lattices with 5-fold and 7-fold rotational symmetries which exist a core state around the inner defect lattices filled with hard-core bosons. First, we consider the FCI/FQAH states with 5-fold rotational symmetry shown in Fig.~\ref{Edge_GPP}. The normal states are marked with blue, while the core state are brown with the angular momentum quantum number 4. If the hard-core bosons do not occupy the core state, based on the GPP and there is one branch of edge excitations with the quasi-degenerate sequence ``1, 1, 2, 3, ...''. If one boson occupy in the core state [with angular momentum quantum number 4 in Fig. ~\ref{Edge_GPP}(a), brown line] and other particles occupy normal states, another branch of edge excitations can be observed with the quasi-degenerate sequence ``1, 1, 2, 3, ...''.  The quasi-degenerate sequence of edge excitations for the FCI/FQAH states with 5-fold rotational symmetry is the total of these two energy branches. For 5 bosons filling in the Chern band, based on the GPP, the root configurations of bosons only occupying normal states are 101010101, 1010101001, 10101010001, 1010100101, 101010100001, 10101001001, 1010010101, etc. The total angular momentum quantum numbers of these root configurations are 20, 21, 22, 22, 23, 23, 23, etc. and the reduced angular momentum quantum number (total angular momentum quantum number modulo $n$) is 0, 1, 2, 2, 3, 3, 3, etc. The root configurations of bosons occupying in one core state are  \underline{\bf 1}01010101, \underline{\bf 1}010101001, \underline{\bf 1}0101010001, \underline{\bf 1}010100101, \underline{\bf 1}01010100001, \underline{\bf 1}0101001001, \underline{\bf 1}010010101, etc. Here, \underline{\bf 1} denotes one hard-core boson occupying the core state. The reduced angular momentum quantum number is 0, 1, 2 ,2, 3, 3, 3, etc. So the quasi-degenerate sequence is ``2, 2, 4, 6, ...''. These occupied root configurations can be shown in Fig.~\ref{Edge_GPP}(a). For 6 bosons filling in the Chern band, the root configurations of bosons only occupying normal states are 10101010101, 101010101001, 1010101010001, 101010100101, 10101010100001, 1010101001001, 101010010101, etc. The total angular momentum quantum numbers of these root configurations are 30, 31, 32, 32, 33, 33, 33, etc. and the reduced angular momentum quantum numbers are 0, 1, 2 ,2, 3, 3, 3, etc. The root configurations of bosons occupying in one core state are  \underline{\bf 1}0101010101, \underline{\bf 1}01010101001, \underline{\bf 1}010101010001, \underline{\bf 1}01010100101, \underline{\bf 1}0101010100001, \underline{\bf 1}010101001001, \underline{\bf 1}010010101, etc. The reduced angular momentum quantum numbers are 4, 0, 1, 1, 2, 2, 2, etc. The total angular momentum quantum numbers of these two energy branches are ``1, 2, 3, 5, ...''.
Then we consider the FCI/FQAH states on singular lattices with 7-fold rotational kagom{\'e} lattice (the core state with with angular momentum quantum numbers 5). Based on the GPP, we can list the root configurations of all bosons occupying in the normal states (shown in Fig.~\ref{Edge_GPP1}) and for 5 bosons filling in Chern band, the reduced angular momentum quantum numbers are 1, 2, 3, 3, 4, 4, 4, etc and for 6 bosons filling in Chern band, the reduced angular momentum quantum numbers are 3, 4, 5, 5, 6, 6, 6, etc. The total quasi-degenerate sequence is ``1, 1, 3, 4, 7, ...''. Similarly, for 6 hard-core bosons filling, the total  quasi-degenerate sequence is ``1, 2, 3, 5, 8, ...''. These exotic edge excitations are different from the 2/5 and 2/3 FQH states which degenerate sequence of edge excitation depends on the number of particles and the angular momentum quantum number of the core state.

\end{document}